\documentclass[11pt,twoside]{article}
\usepackage{asp2014}
\usepackage{graphicx}
\usepackage[usenames,dvipsnames]{xcolor}

\usepackage{hyperref}
\hypersetup{
    final=true,
    pageanchor=true,
    colorlinks=true,
    breaklinks=true,
    linkcolor=blue,
    citecolor=blue,
    urlcolor=blue,
    pdfpagemode=UseNone,
    pdftitle={GREGOR-GRIS Observations},
    pdfauthor={M. Verma},
    pdfsubject={Solar Physics},
    pdfkeywords={Sun: chromosphere, Sun: photosphere, Sun: surface magnetism,
    Sun: sunspots, Methods: data analysis}}

\resetcounters

\begin{document}

\title{Photospheric Magnetic Fields of the Trailing Sunspots in Active Region 
   NOAA~12396}

\author{
    M.\ Verma,$^1$  
    H.\ Balthasar,$^1$   
    C.\ Denker,$^1$ 
    F.\ B\"ohm,$^{1,2}$
    C.E.\ Fischer,$^3$  
    C.\ Kuckein,$^1$
    S.J.\ Gonz{\'a}lez Manrique,$^{1,4}$
    M.\ Sobotka,$^{5}$
    N.\ Bello Gonz{\'a}lez,$^{3}$  
    A.\ Diercke,$^{1,4}$\\ 
    T.\ Berkefeld,$^{3}$
    M.\ Collados,$^{6}$
    A.\ Feller,$^{7}$
    A.\ Hofmann,$^{1}$
    A.\ Lagg,$^{7}$
    H.\ Nicklas,$^{8}$\\ 
    D.\ Orozco Su\'arez,$^{12}$   
    A.\ Pastor Yabar,$^{6, 3}$
    R.\ Rezaei,$^{6}$
    R.\ Schlichenmaier,$^{3}$\\
    D.\ Schmidt,$^{9}$
    W.\ Schmidt,$^{3}$
    M.\ Sigwarth,$^{3}$
    S.K.\ Solanki,$^{7, 10}$
    D.\ Soltau,$^{3}$\\        
    J.\ Staude,$^{1}$
    K.G.\ Strassmeier,$^{1}$
    R.\ Volkmer,$^{3}$
    O.\ von der L{\"u}he,$^{3}$\\ and
    T.\ Waldmann$^{3}$    
\affil{%
    $^1$Leibniz-Institut f{\"u}r Astrophysik Potsdam (AIP),
        Germany; \email{mverma@aip.de}}
\affil{%
    $^2$Humboldt-Universit\"at zu Berlin, Institut fur Physik, Germany}
\affil{%
    $^3$Kiepenheuer-Institut f{\"u}r Sonnenphysik, Germany}
\affil{%
    $^4$Universit{\"a}t Potsdam,
        Institut f{\"u}r Physik und Astronomie, Germany}       
\affil{%
    $^5$Astronomical Institute, Academy of Sciences of the Czech Republic}
\affil{%
    $^6$Instituto de Astrof\'{\i}sica de Canarias, Spain}    
\affil{%
    $^7$Max-Planck-Institut f{\"u}r Sonnensystemforschung, Germany}
\affil{%
    $^8$Institut f\"ur Astrophysik, Georg-August-Universit\"at G\"ottingen,
        Germany}
\affil{%
    $^9$National Solar Observatory, USA}
\affil{%
    $^{10}$School of Space Research, Kyung Hee University, Korea}
\affil{%
    $^{11}$ Departamento de Astrof\'{\i}sica, Universidad de La Laguna, 
         Tenerife, Spain}}  
\affil{%
    $^{12}$Instituto de Astrof\'{\i}sica de Andaluc\'{\i}a, Granada, Spain}

\paperauthor{M.\ Verma}{mverma@aip.de}{0000-0003-1054-766X}{Leibniz-Institut f{\"u}r 
Astrophysik Potsdam (AIP)}{}{}{}{}{}

\begin{abstract}
The solar magnetic field is responsible for all aspects of solar activity. 
Sunspots are the main manifestation of the ensuing solar activity. Combining 
high-resolution and synoptic observations has the ambition to provide a 
comprehensive description of the sunspot growth and decay processes. Active 
region NOAA~12396 emerged on 2015 August 3 and was observed three days later 
with the 1.5-meter GREGOR solar telescope on 2015 August 6. High-resolution 
spectropolarimetric data from the GREGOR Infrared Spectrograph (GRIS) are 
obtained in the photospheric Si\,\textsc{i} $\lambda$ 1082.7 nm and 
Ca\,\textsc{i} $\lambda$1083.9~nm lines, together with the chromospheric 
He\,\textsc{i} $\lambda$1083.0~nm triplet. These near-infrared 
spectropolarimetric observations were complemented by synoptic line-of-sight 
magnetograms and continuum images of the Helioseismic and Magnetic Imager (HMI) 
and EUV images of the Atmospheric Imaging Assembly (AIA) on board the Solar 
Dynamics Observatory (SDO).  
\end{abstract}

\section{Introduction}

\begin{figure*}
\centering
\includegraphics[width=0.72\textwidth]{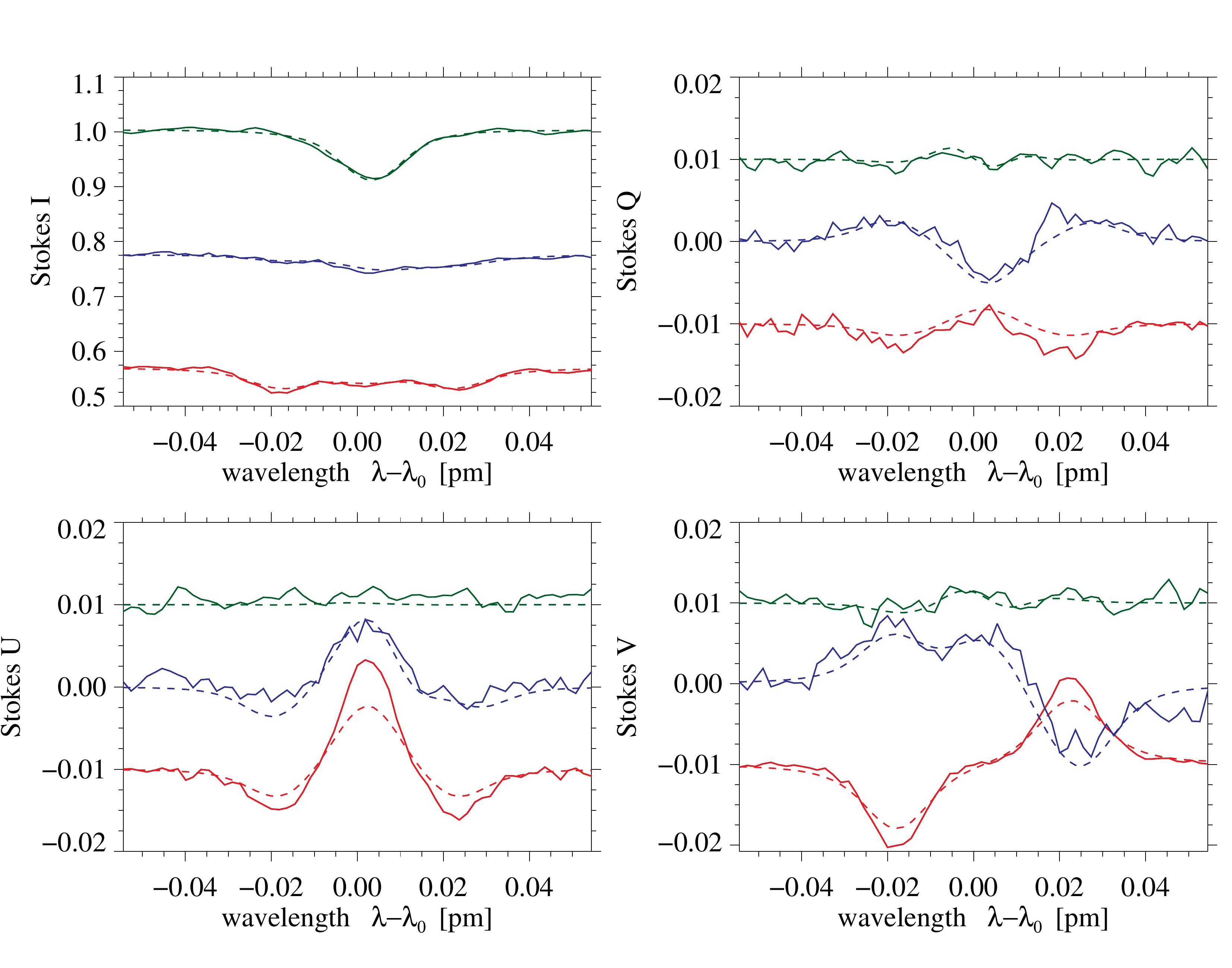}
\caption{All four Stokes-profiles of the photospheric 
Ca\,\textsc{i}~$\lambda$1083.9~nm line for granulation, penumbra, and umbra are 
shown as solid lines in the colors green, blue, and red, respectively. Stokes 
QUV profiles for umbra and granulation were shifted by $-0.01$ and $+0.01$ for 
better visualization. tokes UV profiles for the umbra are scaled down by a 
factor of 2 and 4, respectively, to fit in the displayed range. The SIR profiles 
are shown in the same color-code but as dashed lines.}
\label{FIG01}
\end{figure*}

The present study extends the work by \citet{2016AN....337..1090}, where we 
presented a detailed description of observations and data analysis for one of 
the datasets taken as part of a coordinated observing campaign in August 2015. 
In the initial study, the results from the one-component inversions were 
discussed and were complemented with the SDO observations. Here, we show results 
from two-component inversions. The foremost reason to perform two-component 
inversions was to achieve better fits for double-lobe Stokes V profiles (see 
blue curve in Fig.~\ref{FIG01}). Another motivation was to infer additional 
physical parameters providing more insight into different regimes of magnetic 
fields and with respect to newly emerging magnetic flux.

\section{Observations and Data Analysis}

A two-week coordinated campaign was carried out in August~2015, which included 
observations from the GREGOR, the Vacuum Tower Telescope (VTT), Hinode, and the 
Interface Region Imaging Spectrograph (IRIS). We present one of the datasets 
observed on 2015 August 6, mainly focusing on the spectropolarimetric 
observations from the GREGOR Infrared Spectrograph 
\citep[GRIS,][]{2012AN....333..872C}. The full Stokes polarimetric data were 
taken in the 1083.0~nm spectral range in an 1.8-nm-wide spectral window. This 
window covered the photospheric lines Si\,\textsc{i} $\lambda$1082.7~nm and 
Ca\,\textsc{i} $\lambda$1083.9~nm as well as  the chromospheric He\,\textsc{i} 
$\lambda$1083.0~nm triplet. The 300 scan steps with a step size of 0.144\arcsec\ 
and the image scale of 0.136\arcsec~pixel$^{-1}$ along the slit result in a 
field-of-view (FOV) of about 62\arcsec $\times$ 43\arcsec, which takes about 
30~min to record. The GRIS FOV covers the two trailing sunspots in the active 
region. In this work we mainly discuss the results for the near-infrared 
Ca\,\textsc{i} $\lambda$1083.9~nm line. The Ca\,\textsc{i} line originates from 
the transition between the $4p\ ^3P_2$ and the $3d\ ^3P_2$ levels, and is 
magnetic sensitive with a Land\'e factor $g_\mathrm{eff}=1.5$. 

The basic calibration of the data was carried out on-site using the GRIS 
pipeline \citep{1999ASPC..184....3C}. A careful wavelength calibration was 
performed by comparing the observed profiles with a near-infrared atlas profile 
obtained with the Fourier Transform Spectrometer (FTS) of the McMath-Pierce 
solar telescope at the Kitt Peak National Observatory 
\citep{1996ApJS..106..165W}. We inverted  calibrated GRIS spectra with the, 
Stokes Inversions based on Response functions, (SIR) code developed by 
\citet{1992ApJ...398..375R}. The starting model is a two-component model, which 
covers the optical depth range of $ -4.4 \leq \log \tau \leq +1.0$. We opted for 
the two-component model to improve the fits of complex Stokes V-profiles. 
Examples of the SIR fits in umbra, penumbra, and quiet Sun are shown in 
Fig.~\ref{FIG01}.

\begin{figure*}
\centering
\includegraphics[width=0.73\textwidth]{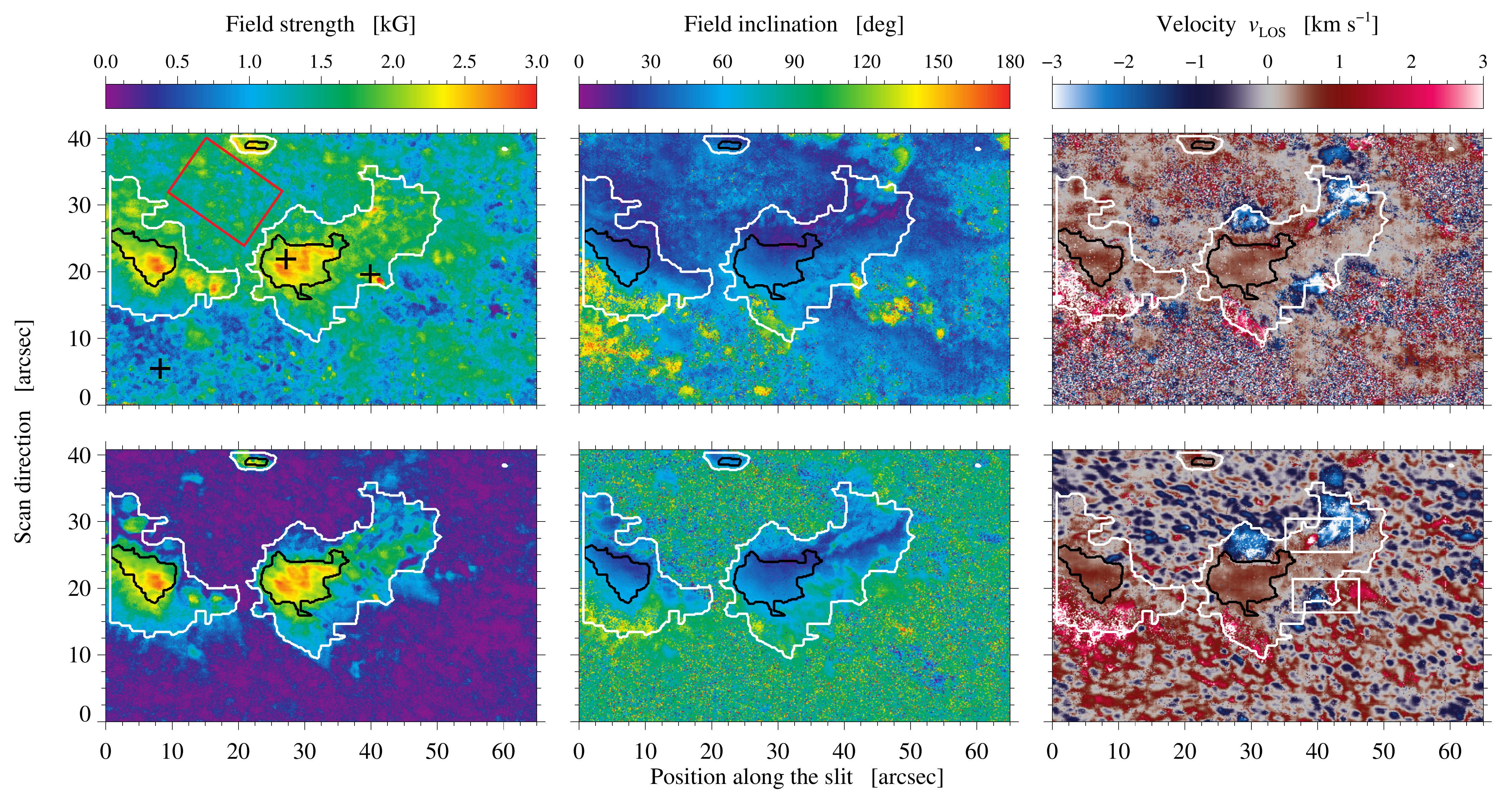}
\caption{Maps of physical parameters derived with the SIR code for the 
Ca\,\textsc{i} line observed by GRIS at 09:24~UT on 2015 August 6: total 
magnetic field strength, magnetic field inclination, and Doppler velocity 
(\textit{left to right}). The two-component inversions (\textit{top and bottom}) 
were sorted according to the filling factor of both magnetic components, and the 
maps corresponding to filling factors larger 0.5 are depicted in the bottom 
panel. The black `+' signs (\textit{top-left}) mark the locations of the 
profiles shown in Fig.~\ref{FIG01} and the red box mark the region with 
kilo-gauss fields. The white boxes (\textit{bottom-right}) label regions with 
red and blueshifts in close vicinity.}
\label{FIG02}
\end{figure*}

\section{Results and Discussions}

Three days after emergence active region NOAA~12396 was still evolving at the 
time of GREGOR observations on 2015 August 6. The SDO/HMI continuum images and 
magnetograms (not shown here) allowed us insight into the temporal evolution of 
the region. They revealed ongoing changes in the appearance of the active 
region's trailing part, especially concerning the evolution of sunspot 
penumbrae. Four hours before the GRIS observations the trailing part consisted 
of three negative polarity sunspots with two sunspots in the south with partial 
penumbra and the one in the north without penumbra. In addition, many mixed 
polarity features were present between the leading and the trailing part. At the 
time of GRIS observations and few hours after we noticed that forming a full 
penumbra is prevented from forming in regions of ongoing flux emergence 
\citep[see][]{2010A&A...512L...1S}.

Full Stokes polarimetry from GRIS provided the photospheric and chromospheric 
line-core intensities, LOS velocities, and magnetic field information for the 
complex active region. Here, we focus on results obtained from the photospheric 
Ca\,\textsc{i} line. The magnetic field information was extracted by conducting 
a two-component inversion with SIR for the Ca\,\textsc{i} line. Two-component 
inversions were carried out because of the inability of one-component inversions 
to fit the double lobes in some of the V-profiles 
\citep[see][]{2016AN....337..1090}. In addition to magnetic field strength the 
inversions provided the line-of-sight (LOS) velocities and magnetic field 
inclination (see Fig.~\ref{FIG02}). We displayed the maps based on the filling 
factor of both magnetic components. The maps corresponding to a filling factor 
larger than 0.5 are less noisy and depict the properties of features like 
sunspot umbrae well. These inversions were able to provide the reasonably good 
estimate of the magnetic field strength with about 2~kG in the umbra and with 
low field inclination. The most conspicuous features in the LOS velocities map 
were the red and blueshifts in close proximity near the penumbra of the right 
spot (see white boxes Fig.~\ref{FIG02}). Matching these location with the HMI 
magnetograms, we deduced that these flows likely belong to regions with flux 
emergence. In the magnetic-field strength maps for low filling factors we found 
kilo-Gauss patches in the area between trailing and leading spot, i.e., the 
region with continuous flux emergence (e.g., red box in Fig.~\ref{FIG02}). These 
kilo-Gauss patches were not visible in the maps based on one-component 
inversions. In addition, the flux emergence in these region seems to inhibit the 
formation of penumbra in the sunspot, which never developed a stable and regular 
penumbra enveloping the whole sunspot. More recently \citet{2013ApJ...771L...3R} 
reported a similar scenario where flux emergence hindered the formation of a 
stable penumbra. 

\acknowledgements The 1.5-meter GREGOR solar telescope was build by a German 
consortium under the leadership of the Kiepenheuer-Institut f\"ur Sonnenphysik 
in Freiburg with the Leibniz-Institut f\"ur Astrophysik Potsdam, the Institut 
f\"ur Astrophysik G\"ottingen, and the Max-Planck-Institut f\"ur 
Sonnensystemforschung in G\"ottingen as partners, and with contributions by the 
Instituto de Astrof\'{\i}sica de Canarias and the Astronomical Institute of the 
Academy of Sciences of the Czech Republic. SDO HMI and AIA data are provided by 
the Joint Science Operations Center -- Science Data Processing. MS is supported 
by the Czech Science Foundation under the grant 14-0338S. This study is 
supported by the European Commission's FP7 Capacities Programme under the Grant 
Agreement number 312495.


\end{document}